\documentclass{aastex}
\usepackage{spr-astr-addons}
\usepackage{url}
	
\RequirePackage{color}

\begin{document}

\title{Timing and emission variation of PSR J1733$-$3716}
\slugcomment{Not to appear in Nonlearned J., 45.}
\shorttitle{Short article title}
\shortauthors{Autors et al.}

\author{Yue Hu\altaffilmark{1,2}} \and \author{Lin Li\altaffilmark{1}}
\and
\author{J.P. Yuan\altaffilmark{2,3}}
\author{S.J. Dang\altaffilmark{2,3}}
\author{S.Q. Wang\altaffilmark{2,3}}
\author{Z.J. Wang\altaffilmark{1}$^\dagger$}
\author{R. Yuen\altaffilmark{2}}
\affil{$^\dagger$Email:wzj@xju.edu.cn}
\altaffiltext{1}{Department of Physics, Xinjiang University, Urumqi, 830046, China.}
\altaffiltext{2}{Xinjiang Astronomical Observatory, Chinese Academy of Sciences, 150, Science 1-street,  Urumqi, Xinjiang 830011, China.}
\altaffiltext{3}{University of Chinese Academy of Sciences, Beijing 100049, China.}

\begin{abstract}
We present analysis of the timing noise in PSR J1733$-$3716, which combines data from Parkes 64-m radio telescope and nearly 15 years of timing data obtained from the Nanshan 25-m radio telescope. The variations in the spin frequency and frequency derivative are determined. The fluctuation in the spin frequency is obvious with an amplitude of 1.94(7)$\times$10$^{-9}$ Hz. Variations of the integrated profile at 1369 MHz are detected with the changes occur in the relative peak intensity from the right profile component. From analysis of the single pulse data at 1382 MHz, we detect weak emission  states that account for 63\% of the whole data, and its duration distribution can be fitted with a power law. The pulsar also exhibits strong emission states, during which the emission shows multiple modes. This includes the normal mode, left mode and the right mode, with the time scales spanning between one and seventeen pulse periods. Such short term variability in pulses contributes to the variation of the integrated profile. Examination of the correlations between the spin parameters and the integrated profiles shows likelihood of a random distribution, which reveals that there is probably no obvious relationship between spin-down rate variations and changes of emission in this pulsar.
\end{abstract}


\section{Introduction}\label{sect:intro}

Pulsars are rapidly rotating and highly magnetized neutron stars, most of which emit pulses regularly by losing
rotational kinetic energy. In spite of the regular rotation, long-term timing has revealed two instabilities in the  pulsar spin-down, namely glitches and timing noise. Glitches are abrupt increases in the rotation frequency, which interrupt the regular spin-down of a pulsar (\citealt{Wang+etal+2000}). Timing noise, which is often presented as non-whiten and un-modeled timing residual, is significant deviation of the pulsar rotation from the spin-down model indicating erratic fluctuation in the rotation of the pulsar. The phenomenon is classified as low-frequency noise or red noise (\citealt{Hobbs+etal+2010}), and usually has time scales of months to years. The cause of timing noise is still unclear, but it has been attributed to either fluctuations in the spin-down or magnetospheric torque (\citealt{cheng1987a, cheng1987b}), or variation of the coupling between the stellar crust and its superfluid core (\citealt{Jones+etal+1990}). It has also been proposed that the macroscopic turbulence in the neutron star’s core superfluid can lead to instability in the spin-down processes (\citealt{melatos2014}). Alternatively, timing noise is thought to arise from small planetary companion (\citealt{Cordes+James+1993}) or free-precession (\citealt{Stairs+Lyne+Shemar+2000}). Despite many interpretations, a full understanding of the mechanism is still lacking.

Although the flux densities and the shapes of single pulses from pulsars are unsteady, it is well known that their integrated pulse profiles are typically stable. However, not all pulsars are consistent with this behavior. 
For dozens of pulsars, their integrated pulse profiles switch between two or more stable shapes. Known as mode changing (\citealt{Backer+1970}; \citealt{Wang+etal+2007}), the phenomenon was first observed in PSR B1237$+$25, which exhibits as switching in the shape of the integrated pulse profile between five-component and three-component (\citealt{Backer+1970}). \citet{Chen+etal+2011} studied mode changing in the strong pulsar B0329$+$54 using Nanshan 25-m radio telescope. They found that the two stable states of this pulsar, namely normal mode and abnormal mode, account for 84.9 percent and 15.1 percent of the total duration, respectively. In the investigation of the young pulsar PSR B1828$-$11, it was revealed that the pulsar displays two stable modes, the wide mode and the narrow mode, with an estimation of around 500 days for the period of change between the two modes (\citealt{Stairs+etal+2019}). Another phenomenon relates to pulse nulling, which could be regarded as an extreme mode changing, and demonstrates as disappearance of pulses for durations ranging from several pulse periods to many years (\citealt{Wang+etal+2007}). Pulse nulling is generally considered as due either to the null state or the weak state, from which pulses are not detected because their flux densities drop below the sensitivity of the telescope.

Observations have shown that changes in the pulse profile in a few pulsars are associated with the instabilities in the rotation of the stars. \citet{Lyne+etal+2010} 
found that some pulsars with quasi-periodicity in the timing noise show switching between two different spin-down rates. Furthermore, the profile width of six pulsars were identified to correlate with the spin-down rate. Both PSRs B0919+06 and B1859+07 exhibit spin-down state switching, and the changes in the spin-down rate are modulated with the variations in the pulse shape (\citealt{Perera+etal+2015}; \citealt{Perera+etal+2016}). From the application of Gaussian process regression to 168 pulsars, \citet{Brook+etal+2016} discovered a strong correlation between pulse shape changes in PSR J1602$-$5100 and its spin-down rate. Following a glitch event, PSR J0742$-$2822 was detected with stronger correlation between its spin-down and pulse shape (\citealt{Keith+etal+2013}). PSR B2035+36 switches between two emission modes after a glitch, which is also accompanied by a narrower pulse shape (\citealt{Kou+etal+2018}). It was suggested that both emission change and variation in the pulsar rotation are caused by changes in the flow of the magnetospheric current (\citealt{kramer+2006}). Therefore, more investigation on variations in the emission and in the pulsar rotation may shed new light on the physical processes in pulsar magnetosphere.

PSR J1733$-$3716 was discovered in the high frequency pulsar survey in southern Galactic plane (\citealt{Johnston+etal+1992}). It is a normal pulsar with a rotation period of 338 ms and a dispersion measure (DM) of 153.18 $\rm pc~cm^{-3}$ 
(\citealt{Petroff+etal+2013}). This pulsar has a period derivative of 1.5$\times$10$^{-14}~$s$^{-2}$, and characteristic age of 355 ky. Its integrated pulse profile has two peaks with the left peak being higher in intensity. Proper motion has been detected with value of $\mu_{\alpha}=4(9)$ mas yr$^{-1}$ and $\mu_{\delta}=63(34)$ mas yr$^{-1}$ (\citealt{Li+etal+2016}). Recent work on PSR J1733$-$3716 has updated its position and rotation frequency parameters (\citealt{Parthasarathy+etal+2019}). In this paper, we study the timing behavior and emission variation of PSR J1733$-$3716. In Section 2, we describe the observing systems and the data reduction process. In section 3, we report the results of timing analysis, and changes in the shape of the integrated pulse profile and  in the single pulses 
are presented. We discuss our results and conclude the paper in Section 4.

\section{Observation and Data Analysis}
\label{sect:Obs}

\subsection{Timing data and analysis}
\label{sect:timing and analysis}

The regular timing observations of PSR J1733$-$3716 have been carried out with Nanshan 25-m radio telescope since July 2002. A dual-channel cryogenic receiver is used, which operates from 1380~MHz to 1700~MHz with a bandwidth of 320~MHz and a center frequency of 1540~MHz. Signals are recorded by an analogue filter bank (AFB) with 2$\times$128$\times$2.5~MHz channels (\citealt{wang+etal+2001}). As from January 2010, a digital filter bank (DFB) with 2$\times$1024$\times$0.5~MHz channels has come into operation. The observing cadence is about three times per month with each observation lasts for 16\,min. Due to antenna maintenance, there was a cut-off time in the Nanshan data set on MJD 57150.

The timing data for PSR J1733$-$3716 were also obtained with Parkes 64-m radio telescope at frequencies centered at 1369 MHz, 1374 MHz, 1465 MHz and 3094 MHz. The data\footnote{\url{https://data.csiro.au/dap/public/atnf/pulsarSearch.zul}} were recorded by the analogue filter bank (AFB) and one of the Parkes Digital Filter-Bank systems (PDFB1/2/3/4). The observations that used PDFB were folded online using 8-bit digitization to provide a typical 512 bins per period and 1 min sub-integration intervals. Information of all the timing data used in this paper is shown in Table \ref{timobs}.

In order to determine the pulse times of arrival (ToAs), we employed the \texttt{\textsc{psrchive}} software package (\citealt{Hotan+etal+2004}) to conduct the offline processing. The steps are as follows:

i) Use of the \texttt{\textsc{psrchive}}'s \texttt{pazi} tool to remove the radio frequency interference, followed by summing the data in frequency, time and polarization to produce an integrated pulse profile for each observation. Here, the timing models are obtained from ATNF Pulsar Database\footnote{\url{https://www.atnf.csiro.au/research/pulsar/psrcat/expert.html}}.

ii) Application of the \texttt{\textsc{psrchive}}'s \texttt{pat} tool to determine
 pulse-of-arrival by cross-correlating the integrated profiles with a template, and the latter was obtained by summing all available data for each system. We then employ the \textsc{EFACEQUAD} plugin to estimate the excess scatter of white noise in the residuals. This plugin introduces two parameters, which are the EFAC and EQUAD, such that the original uncertainty in the ToAs, $e$, is converted to the new $e'$ with $e'^2=\rm EFAC^2\times(e^2+\rm EQUAD^2)$.

iii) Removal of the effects due to the Earth's motion by correcting the ToAs to the solar-system barycenter using the solar-system ephemeris DE421 (\citealt{Folkner+etal+2009}) and the time system Barycentric Coordinate Time. We then employed the standard timing program software package \texttt{\textsc{tempo2}} (\citealt{Edwards+Hobbs+Manchester+2006}; \citealt{Hobbs+Edwards+Manchester+2006}) to obtain a timing solution, using the least-squares fitting. Offset between different observing systems were also included.

iv) Following \citet{Li+etal+2016}, we analyze the proper motion and timing noise using the frequentist method, and the Cholesky decomposition of the covariance matrix is used to describe the stochastic processes in the ToA data.

\begin{table}[t]
\begin{minipage}[]{80mm}
\caption[]{Information of the timing observations.
\label{timobs}}\end{minipage}
\setlength{\tabcolsep}{3pt}
\footnotesize
 \begin{tabular}{lll}
  \hline\noalign{\smallskip}
Telescope & Nanshan & Parkes 					\\
   \hline\noalign{\smallskip}
Data span (MJD) & 52500$-$57140 & 52001$-$57705 \\
Numbers of ToAs & 268           & 139           \\
Central Freq (MHz) & 1540 & 1369;1374;1465;3094 \\
Backend &  AFB,PDFB &   AFB,PDFB   						\\
Integration time(min) & 16 & 3;6;10 			\\
  \noalign{\smallskip}\hline
\end{tabular}
\end{table}

\begin{figure}
   \centering
   \includegraphics[width=8cm,angle=0]{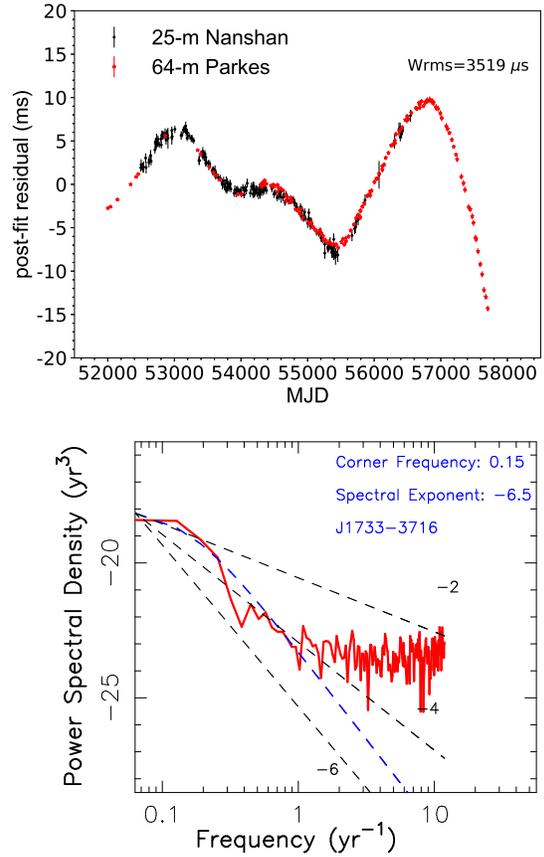}\\
  \includegraphics[width=5.5cm,angle=-90]{huyue_fig1b.eps}
   \vspace{5mm}
   \caption{Top: The timing residuals of PSR J1733$-$3716, with position,
   spin frequency and frequency derivative fitted. The circles and asterisks in the plot
   represent Nanshan and Parkes data, respectively. Bottom: Power spectrum of the timing noise.}
   \label{timres}
\end{figure}

\begin{figure}
   \centering
   \includegraphics[width=8.0cm,angle=0]{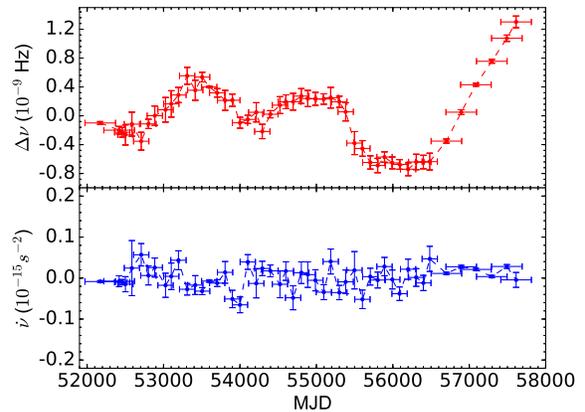}
   \caption{The upper panel shows the residual frequency, with a spin-down
   mode subtracted. The lower panel displays the change of spin frequency
   derivative.}%
   \label{nunudot}
   \end{figure}

\subsection{Single-pulse data and analysis}

\begin{table}[t]
\begin{minipage}[]{82mm}
\caption[]{The information of single pulse observations for\\ PSR J1733$-$3716 using Parkes 64-m telescope  at center frequency of 1382MHz.
  \label{sglobs}}
\end{minipage}
\setlength{\tabcolsep}{5pt}
\footnotesize
\begin{tabular}{cccccc}  
  \hline\noalign {\smallskip}
Num & Date & MJD & S/N & Length(s) & Pulse-Num\\
 \hline\noalign {\smallskip}
1 & 20140728 & 56867 & 59.135 & 186.662 & 553\\
2 & 20140820 & 56890 & 52.604 & 186.855 & 554\\
3 & 20150417 & 57130 & 52.321 & 186.885 & 554\\
4 & 20150523 & 57166 & 57.089 & 189.001 & 560\\
5 & 20150620 & 57194 & 62.318 & 186.785 & 554\\
6 & 20150725 & 57229 & 58.992 & 191.604 & 568\\
7 & 20150811 & 57246 & 58.115 & 195.014 & 578\\
8 & 20151014 & 57309 & 48.638 & 186.887 & 554\\
9 & 20170605 & 57910 & 59.387 & 187.846 & 557\\
10 & 20170701 & 57936 & 41.409 & 188.844 & 560\\
11 & 20170827 & 57993 & 44.115 & 191.057 & 566\\
  \hline\noalign{\smallskip}
   &        &       & SUM  & 2077. 44   & 6158 \\
  \noalign{\smallskip}\hline
\end{tabular}
\end{table}

The single pulse observations of PSR J1733$-$3716 were performed using the Parkes 64-m radio telescope at center frequency of 1382 MHz and bandwidth of 400 MHz. The data were obtained between MJD 56866 and MJD 57993 (July 2014 and August 2017) with the Berkeley Parkes Swinburne Recorder (BPSR), which has 1024 frequency channels. The signals were sampled every 64 $\mu$s for a total of 2077 s (or equivalent to 6158 pulse periods). The details of the observations are shown in Table \ref{sglobs}. Steps for data processing are as follows:

i) With the best-fitting ephemeris obtained from \texttt{\textsc{tempo2}}, we extracted individual pulses with 1024 phase bins per pulse period using the \texttt{dspsr} program (\citealt{Van+Bailes+2011}).

ii) Radio frequency interference (RFI) was removed automatically using \texttt{paz} tool in the \texttt{\textsc{psrchive}} software package, and the remaining RFI was interactively removed using \texttt{pazi} tool.

iii) Data were summed in frequency and polarization using \texttt{pam} in the \texttt{\textsc{psrchive}}. At the end, we obtained 6158 single pulses in total.

\section{Result}
\label{sect:reduction and result}
\subsection{Timing noise and variations of the spin parameters}
\label{sect:Timing}

\citet{Coles+etal+2011} developed a method for determining pulsar timing solution by applying the Cholesky decomposition  on the covariance matrix of timing residual. They suggested that analyzing the power spectrum of the timing noise is the optimal technique to characterize strong red noise. Based on the initial ephemeris obtained from PSRCAT, we fit the timing model incorporating the frequentist method which provides an unbiased least-squares fit even in the presence of red noise (\citealt{Coles+etal+2011}). The post-fit timing residuals for PSR J1733$-$3716 with fitted rotational frequency, frequency derivative and position are illustrated in Fig. \ref{timres}. The position at epoch MJD 55438 is obtained with RAJ = 17:33:26.7624(13) and DECJ = $-$37:16:55.23(6),  and the epoch of MJD 55438 is set on the middle of the whole data span. By comparing with the earliest position reported by \citet{Johnston+etal+1995}, 
we obtain the proper motion with $\mu _{\alpha }^{\rm {F}}$=$-$3(5) mas yr$^{-1}$ and $\mu _{\delta }^{\rm {F}}$=40(18) mas yr$^{-1}$. Table \ref{timpara} lists the updated parameters for PSR J1733$-$3716, including its position, rotational frequency, frequency derivative and proper motion, which are consistent with previous measurements (\citealt{Li+etal+2016}; \citealt{Parthasarathy+etal+2019}).

The bottom plot in Fig. \ref{timres} shows the power spectrum for the timing noise in PSR J1733$-$3716. It is evident from the plot that the red noise  dominates the stochastic process. The power spectrum of timing noise in pulsars can be  modeled using a power-law $P(f)=A/[1+(f/f_{\rm c})^2]^{\alpha/2}$, where $A$, $f_{\rm c}$, and $\alpha$ represent amplitude, corner frequency and spectral index, respectively (\citealt{Coles+etal+2011}). A mode with $\alpha=-6.5$ and $f_{\rm c}=0.15$ yr $^{-1}$ could nicely describe the power spectrum of the timing noise in PSR J1733$-$3716.

\begin{table}[t]
\begin{minipage}[]{82mm}
\caption[]{Parameters of PSR J1733$-$3716 obtained using \\the frequentist method.
\label{timpara}}\end{minipage}
\setlength{\tabcolsep}{5pt}
\footnotesize
 \begin{tabular}{lcc}
  \hline\noalign{\smallskip}
               & \citet{Li+etal+2016}  & This work   \\
 \hline\noalign{\smallskip}
  RAJ (h:m:s)	& 17:33:26.7628(3)& 17:33:26.7624(13) \\
  DECJ (d:m:s) & $-$37:16:55.23(2)& $-$37:16:55.23(6)\\
  EPOCH (MJD)  & 54433            & 55438\\
  $\nu$ ($\rm s^{-1}$)  & 2.9621405983(10) & 2.9621291821(11) \\
  $\dot{\nu }$ ($\rm 10^{-14}~s^{-2}$) &$-$13.20051(18)  & $-$13.20044(19) \\
  $\mu _{\alpha }^{\rm {F}}$ ($\rm mas\, yr^{-1}$)& 4(9) & $-$3(5)\\
  $\mu _{\delta }^{\rm {F}}$ ($\rm mas\,yr^{-1}$) & 63(34) & 40(18)\\
  Time scale & \multicolumn{2}{c}{TCB}\\
  Ephemeris &  \multicolumn{2}{c}{DE421}\\
 \hline\noalign{\smallskip}
\end{tabular}

\end{table}

\begin{figure}
   \centering
   \includegraphics[width=8.1cm,angle=0]{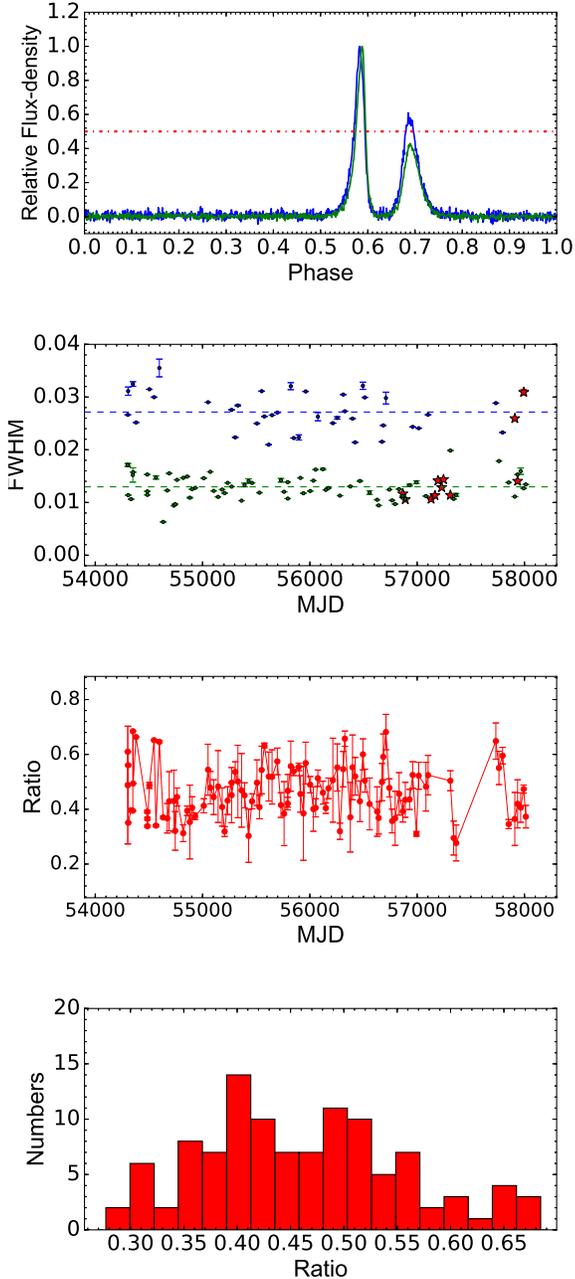}
   \vspace{-10mm}
   \caption{Top two panels: The integrated profiles and the corresponding FWHM of PSR J1733$-$3716.
 The blue color represents the high state and the green color represents the low state  at 1369 MHz.  The red pentagram represents the FWHM of integrated profiles in 1382 MHz. The blue and green dash lines represent the average FWHMs  for the ``two states'', which are 0.027(4) and 0.013(2) phase, respectively.
   Bottom two panels:  Intensity ratio of the right profile peak to the left profile peak and its statistical distribution histogram for PSR J1733$-$3716. }
   \label{w50}
\end{figure}

The variation of spin parameters with time is examined. The values of $\nu$ and $\dot\nu$ were determined from fitting  the spin-down model to data sections, each has duration of 200 days with an overlapping of 100 days between adjacent sections. After MJD 56500, infrequent observations forced us to set the length of a section to typically 400 days, with an overlapping of 200 days between adjacent sections.
We obtained a series of values, which are plotted in Fig. \ref{nunudot}, with uncertainty of 3$\sigma$ (that is, triple the standard used for fitting in TEMPO2). Here we use the error-bars in the x-direction to show the extent of each fit window. Perturbation of residual in the rotational frequency is clearly seen in the figure with peak-to-peak amplitude of $\sim$1.94(7)$\times$10$^{-9}$ Hz, which corresponds to a relative variation of $0.65(2)\times10^{-9}$. The residual frequency decreases by $\sim$ 9.4(6)$\times$10$^{-10}$ Hz in 450 days between MJD 55250 -- 55700, whereas it increases by $\sim$ 1.94(7)$\times$10$^{-9}$ Hz more gradually over 1400 days between MJD 56300 -- 57700. The spin-down rate also fluctuates, with peak-to-peak amplitude of 1.2(3)$\times$10$^{-16}$ s$^{-2}$  and fractional variation $\Delta\dot\nu$/$\dot\nu$ of 0.09(3)\%. Overall, the fluctuation in spin-down rate is smooth. The bottom  panel in Fig. \ref{nunudot} shows that the spin-down rate between MJD 56500 -- 57705 is decreasing, which correlates with the rising trend in the residual frequency during this period.

\subsection{Apparent variation of the integrated profile}
As the integrated pulse profile  of PSR J1733$-$3716 has  two peaks (\citealt{Johnston+etal+1995}), whose shapes may change, we investigate variation of the pulse profile.
The second plot from the top in Fig. \ref{w50} shows the full width at half maximum (FWHM) of the integrated pulse profiles normalized at the peak intensity at 1369 MHz. It clearly displays a bimodal distribution, which indicates that FWHM switches between ``two states'', referred to as the high state, which has an average  FWHM of 0.027(4), and the low state, with an average FWHM of 0.013(2). The number of points for the high state is less than that for the low state, with 31.6\% of  the time this pulsar is in the high state. We summed all the  pulses in the high state and in the low state separately to obtain normalized integrated profiles, which are shown in the top panel in Fig. \ref{w50}.
Both pulse profiles show a double-peak structure, with each displaying similar shape of Gaussian function. In addition, we identify variability in the right profile peak.
The relative intensity of the right profile peak in the high state is $\sim$0.57, which is more than half of the left peak intensity, whereas it is $\sim$0.41 in the low state, less than half of the left peak intensity.

We plot the ratio of the two profile peak intensities for each observation and its statistical distribution, which are shown in the two bottom panels in Fig. \ref{w50}. The average  ratio of the right peak to left peak is 0.46(9). It is clear that the ratio changes randomly with time and the corresponding histogram does not show obvious bimodal distribution. 
Besides, Chopping each 1369 MHz observation in half, we obtained the FWHMs of each half.
It was given that, both halves of 55 over 106 observations are in the “same state”.
The other 51 observations show different “state” between halves,
which is caused by rapid variation of bright pulses  and
short integration time for each observation (see section \ref{single-pulse}).
We conclude that the apparent variations of the integrated profile are neither mode change  nor real long-term changes.

\subsection{Two states of single pulses}
\label{single-pulse}

\begin{figure}
   \centering
 \includegraphics[width=8cm,angle=0]{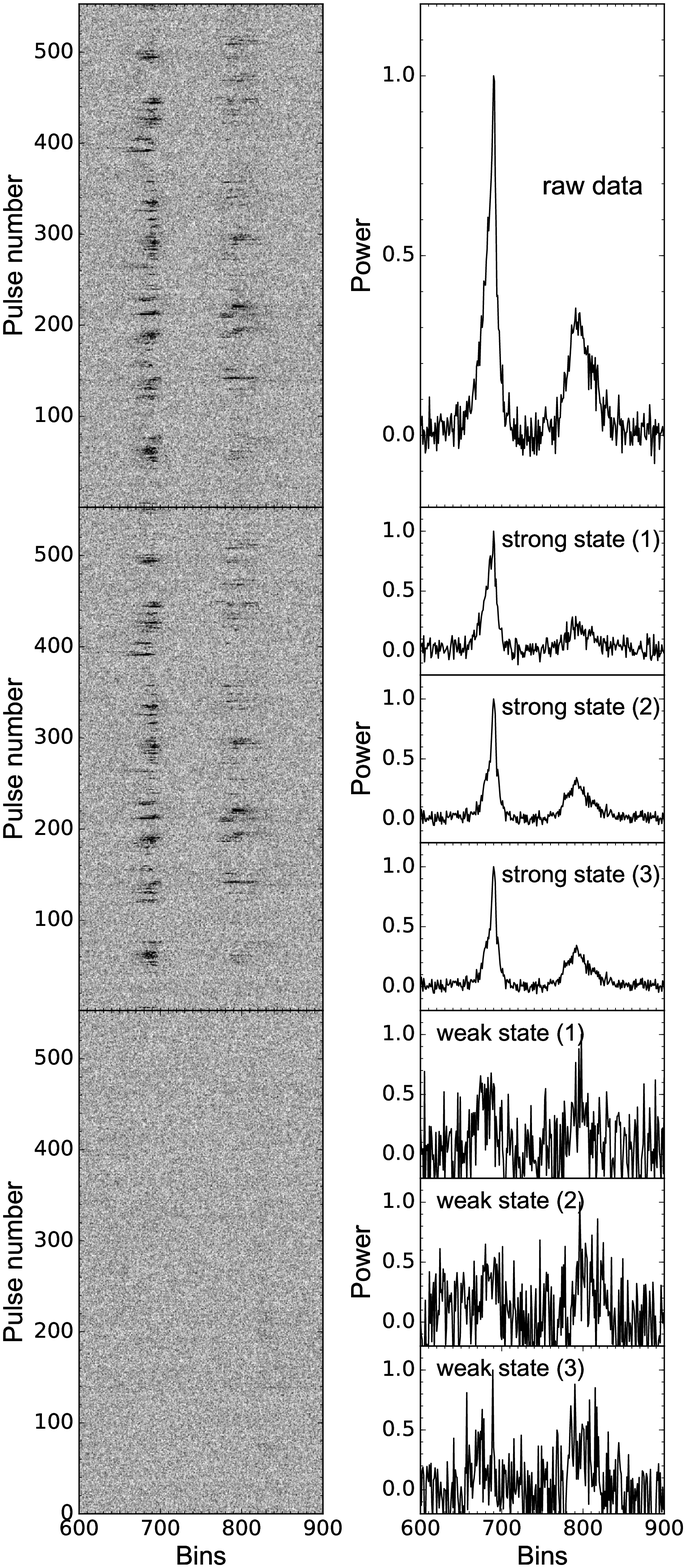}
 \caption{Single pulse sequence obtained on MJD 56867 is shown in grayscale in the three panels on the left hand column. From top to bottom, the pulses are obtained from the raw data, strong state (in which pulses from the weak state are replaced by white noise) and weak state  where pulses from the strong state are replaced by white noise). The panels on the right hand column show the integrated pulse profiles, obtained using, from top to bottom, the raw data, strong state (divided into three sections) and the weak state (divided into three sections). All the integrated pulse profiles are normalized.}
\label{weak}
\end{figure}

    \begin{figure}
   \centering
 \includegraphics[width=8cm,angle=0]{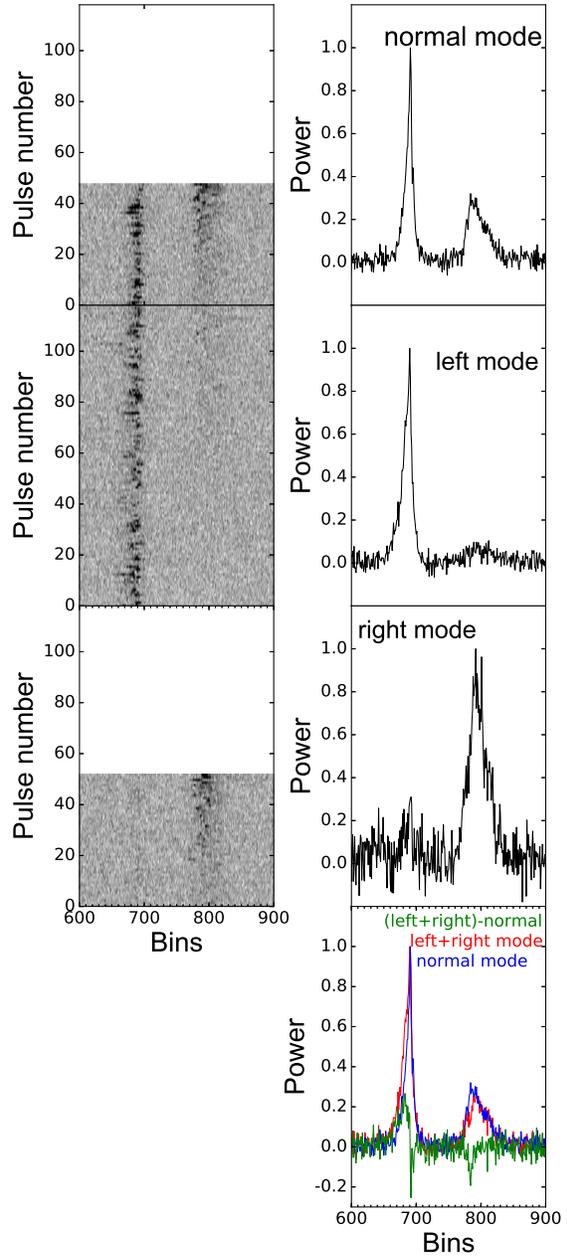}
 \caption{ Left column: Sequences of selected single pulses are presented to illustrate the normal mode (top panel), left mode (middle panel) and the right mode (bottom panel). The corresponding integrated  pulse profiles for the three different modes are shown in the right hand column. In the  bottom right panel, the red solid line represents the profile obtained from summing the single pulses in left and right modes. The green solid line signifies the difference between  this `summed' profile and the normal mode profile. }
   \label{3modes}
   \end{figure}

 \begin{table*}
 \centering
  \caption[]{Statistics for the weak and strong states in all 11 sections of data. The strong state consists of three different modes, which are the normal, left and right modes.
\label{proportion}}
\setlength{\tabcolsep}{2pt}
\footnotesize
 \begin{tabular}{cccccccl}
  \hline\noalign{\smallskip}
Num & MJD & Weak state~ & Strong state~ & Normal mode~ &Left mode~  & Right mode~ \\
     &     & (\%) & (\%) & (\%) & (\%) & (\%)\\
 \hline\noalign{\smallskip} 
1 & 56867 & 60.58 & 39.42 & 8.68 & 21.34 & 9.40 &\\
2 & 56890 & 62.82 & 37.18 & 9.39 & 16.60 & 11.19 &\\
3 & 57130 & 63.00 & 37.00 & 5.96 & 18.23 & 12.81 &\\
4 & 57166 & 61.96 & 38.04 & 6.61 & 16.61 & 14.82 & \\
5 & 57194 & 57.76 & 42.24 & 11.55 & 19.86 & 10.83 &\\
6 & 57229 & 60.74 & 39.26 & 11.09 & 16.73 & 11.44 &\\
7 & 57246 & 63.15 & 36.85 & 7.43 & 18.17 & 11.25 &\\
8 & 57309 & 60.82 & 39.18 & 9.19 & 16.32 & 13.67 &\\
9 & 57910 & 65.88 & 34.12 & 7.90 & 16.70 & 9.52 &\\
10 & 57936 & 69.29 & 30.71 & 5.00 & 18.21 & 7.50 &\\
11 & 57993 & 69.79 & 30.21 & 7.24 & 13.25 & 9.72 &\\
  \noalign{\smallskip}\hline
\end{tabular}
\end{table*}


A contiguous sequence of single pulses from PSR J1733$-$3716 obtained on 28 July 2014 is shown in the three panels on the left hand column in Fig. \ref{weak}. It demonstrates frequent and complicated variations in the pulse profile. Instances of apparent nulling can also be seen. In order to distinguish different states,
we use the method proposed by \citet{Bhattacharyya+etal+2010} to identify  a radio state in which significant pulses from the pulsar are detected. In this process, we conducted two operations: (1) sorting out single pulses based on their energies, and (2) using 3$\times\sqrt{n_{\rm on}}\sigma_{\rm off}$ as the threshold to classify pulses with ON pulse energy smaller than it as apparent null, where $n_{\rm on}$ is the number of ON pulse bins and $\sigma_{\rm off}$ is the rms of the OFF pulse region.
The results obtained from data labeled NUM 1 are shown in Fig. \ref{weak}. Two different states are identified, and their  corresponding normalized integrated profiles are shown in  the panels on the right hand column in Fig. \ref{weak}. It can be seen that the apparent nulling is different from white noise, which is an indication of weak emission and we define this as weak state. It is clear that the SNR of the strong state is higher than that of the original data. 

From the investigation of the single pulses in the strong state, we find that there are at least three different radio modes: the left mode (radiation from left profile component only), right mode (radiation from right profile component only) and the normal mode (both left and right profile components radiate). Using similar method, we identify the left mode, in which the pulses from the right profile component become null, the right mode, in which pulses forming the left profile component disappear. The other active pulses are in the normal mode. Fig. \ref{3modes} shows the three different radio modes, where only corresponding pulses are selected and presented.
For integrated profile in the normal mode, the right profile component has a peak intensity that is 0.3 times that from the left. As the duration of the pulses in left mode accounts for half of the strong state, the left mode dominates in the strong state. For integrated profile in the left mode (shown in bottom panel of Fig. \ref{3modes}),
we find that it exists 
 a weak profile in the region of right profile component. 
 It implies that, for the left mode, radiation from the right profile component is in weak state instead of a null. By comparing the pulse profile obtained from summing the left and right  modes to that obtained from the normal mode, we identify obvious differences between them, which is shown in the bottom panel of Fig.  \ref{3modes}.  Therefore, we infer that the normal mode is a separate mode,  and not merely a combination of the left and right modes.

\begin{figure}
   \centering
   \includegraphics[width=8cm,angle=0]{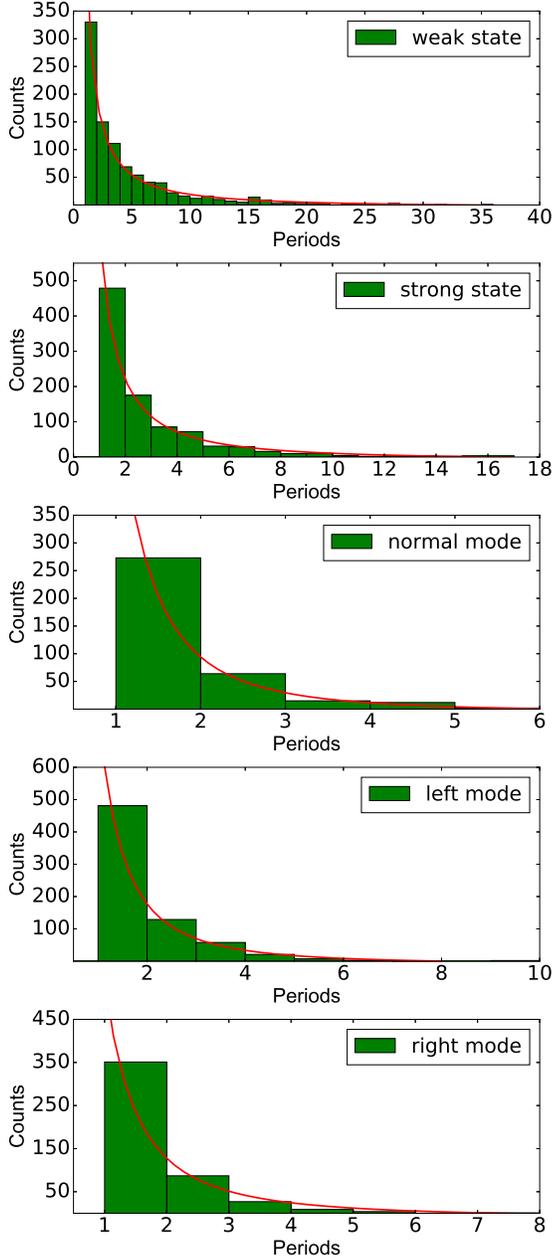}
   \caption{Histogram for the durations of the weak state, strong state, normal mode, left mode and the right mode. The red curve in each plot illustrates the modeled distributions using power-law.}
   \label{countdura}
   \end{figure}

We count the duration for the two states in each of  the 11 data segments. Table. \ref{proportion}
shows the proportion  for the different states, of which the weak state accounts for 63.45\% suggesting that the weak state dominates in each data  segment. During the strong state, 47.81\%, 29.76\% and 22.43\% of the duration are in the left mode, right mode and normal mode, respectively. 
 The histogram of durations  for all identified weak states are shown in Fig. \ref{countdura}. It can be seen that the  duration of weak state  ranges from 1 to 36   pulse periods (or 0.34 -- 12.15 s). The distribution of durations can be fitted by a power law with a slope of $-1.3$. For  the strong state, the duration is found to range from 1 to 17 pulse periods (or 0.34 -- 5.74 s), with one period
dominates the distribution as shown in Fig. \ref{countdura}. The distribution of durations for the strong state can be fitted by  a power law which has a slope of $-1.5$. For  the normal mode, left mode and  the right mode, their durations are
found to range from 1 to 10  pulse periods  (or 0.34 -- 3.38 s), with one period dominates the distribution.
The distribution of durations for the normal mode, left mode and  the right mode can also be fitted by power law with a slope of $-2.6$, $-2.1$ and $-2$, respectively.
Therefore, PSR 1733$-$3716  switches between two states frequently, with  the weak state lasts for 0.34 -- 12.15 s, and strong state lasts for 0.34 -- 5.74 s. Furthermore, the durations for the three modes are shorter in strong state, with  the normal, left and right modes last for 0.34 -- 1.69 s, 0.34 -- 3.04 s, and 0.34 -- 2.36 s, respectively.

\section{Discussion}

By analyzing more than 15 years of timing observations on PSR J1733$-$3716, we identify significant fluctuations in the spin frequency of this pulsar with typical red noise. For PSR J1733$-$3716, the fractional amplitude of variation in the spin frequency 
($\Delta\nu/\nu \sim 0.65(7)\times10^{-9}$ Hz) is similar to that of small glitches. The power spectrum of red noise can be well modeled by a power law with a spectral index of $-6.5$, which is close to $-6$. This implies that random walk probably dominates the timing noise in  the spin-down rate, although timing noise could not be explained as a pure random walk (\citealt{Hobbs+etal+2010}).

The integrated profile of PSR J1733$-$3716 is characterized by two-component shape.
We detected apparent variation in the integrated profile of this pulsar, that is, FWHM switches between ``two states'', resulting  in the fluctuation  in the relative intensity  from the right  profile component. In fact, this is caused by rapid variations of  bright single pulses, which lead to many short-term variations  in the relative  intensity in the right  profile component due to several minutes integration time. Therefore, the apparent variations of the integrated profile are not truly long-term changes.  This is different from PSR B2035+36, with which an actual variation occurs causing the emission to switch between two states (\citealt{Kou+etal+2018}). The causes of the pulse profile variability include spin-precession, phase-shifting or ``flare'', change of inclination angle, inaccurate DM value, scatter broadening or intrinsic emission changes (\citealt{Rankin+etal+2006}; \citealt{Brook+etal+2018}; \citealt{Desvignes+etal+2019}). As for PSR J1733$-$3716,  the intrinsic variation of emission inherited in each single pulse  may contribute to the variability of the integrated profile.

\begin{figure}[t]
   \centering
   \includegraphics[width=8.0cm,angle=0]{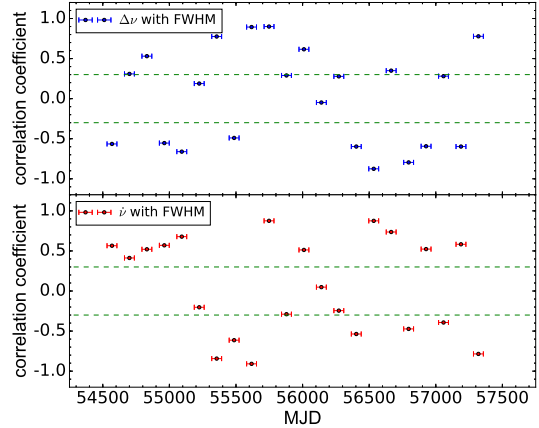}
   \caption{Top panel:  correlation coefficients between FWHM and residual frequency $\Delta\nu$. Bottom panel:  correlation coefficients between FWHM and the derivative of spin frequency $\dot{\nu}$. The two dash lines in each plot represent the values of 0.3 and $-0.3$.}
   \label{correlation}
   \end{figure}

Variations  in the integrated  profile have been detected to correlate with rotational changes in 
pulsars (e.g \citealt{Brook+etal+2019}). Apart from intermittent pulsars, there are ten radio pulsars which show connections between spin-down rate variation and change of emission, which suggest that pulsar rotation is correlated with emission for some pulsars. A  reduce in the total profile intensity, would signify an overall drop in the plasma density (e.g., \citealt{Taylor+Manchester+1975}), which in turn imply a decrease in the current flow leading to a decrease in the spin-down rate.
We search  for the above relationship in PSR J1733$-$3716 by calculating the correlation coefficients between FWHM and frequency residual, and between FWHM and the first derivative of the spin  frequency. The results are presented in Fig. \ref{correlation}, which shows that the correlation coefficients are likely random. Therefore, we infer that there is probably no obvious relationship between variation of spin-down rate and change of emission in PSR J1733$-$3716.

The short-term emission variability in this pulsar is detected by detailed analysis of single pulse data using the Parkes 64-m radio telescope. A total of two states, namely a weak state and a strong state, are determined. The weak state is detected with fraction of about 63\%, indicating that the weak state  dominates this pulsar. The existence of the weak state and three distinct radio modes in the strong state as unveiled by the
single-pulse observations reveals that variations of the emission properties in this pulsar are unique at different parts of the profile. The assumption that pulsar emission is generated by outflow of relativistic pair plasma along open field lines in two-stream instability implies that pulsar radio flux density varies in proportional to the plasma density (\citealt{Taylor+Manchester+1975}; \citealt{Cordes+1979}; \citealt{Lyubarskii+1996}). Furthermore, recent observations indicate that pulsar magnetospheres can exhibit multiple states of distinct emission properties (\citealt{Smits+etal+2005}; \citealt{kramer+2006}; \citealt{Rankin+etal+2013}; \citealt{Basu+Mitra+2018}) implying variations in the emission region. For example, PSR B1918+19 shows coexistence of nulling and multiple modes of subpulse drifting across the profile window (\citealt{Rankin+etal+2013}), and PSR B1931+24 exhibits two states of radio emission each associated with a unique charge density in the plasma (\citealt{kramer+2006}). In the case of PSR J1733$-$3716, a unique profile shape in different radio modes suggests that each mode corresponds to unique emission properties in the emission region. In addition, the disappearance of a different profile component in different modes indicates that the plasma properties change uniquely at different parts of the profile between different modes. When expressed in the magnetic frame, the emission properties change as functions of the polar, $\theta_b$, and azimuthal, $\phi_b$, angles, and both are function of rotational phase, $\psi$, of a pulsar. For dipolar structure, a field line is specified by two constants namely $r_0 = r/\sin^2 \theta_b$, which is the field line constant, where $r$ is the radial distance from stellar center, and $\phi_b$, which signifies the azimuthal position of the field line at the surface of the star. This allows the unique emission properties associated with the plasma in the observable emission region for each mode to be specified by $(\theta_b,\phi_b)$ through modeling of the changes at different parts of the profile. This will offer more in-depth investigation of the emission mechanism underlying a broader pulsar phenomena that show similar behavior, such as mode-changing and nulling. To do that would require a more sophisticated description of this version of pulsar radio emission which involve detailed modeling on the plasma processes in distorted field line structure, and we plan to discuss it in the future.

\section{Summary}
We have presented the timing results for radio pulsar PSR J1733$-$3716,  in which the timing residuals are dominated by red noise. The power spectrum of  the timing noise in PSR J1733$-$3716 is obtained, as well as accurate  values for the rotation frequency parameters, position and proper motion. We then analyze the variation in the spin parameters. Both frequency and frequency derivative are found to fluctuate with time. By calculating FWHM of the integrated profile, we find that the integrated profile is not stable but demonstrates variation in the relative intensity at the peak of the right profile component in each observing session.
From the investigation of the single-pulse data of PSR J1733$-$3716, we detect two different states, which are the strong state and the weak state, with the weak state dominating 63.45\%  of the total duration. In addition, multiple modes, namely normal mode, left mode and right mode, are revealed and they have time scales ranging from one to seventeen pulsar periods.

Finally, we  calculate the correlation coefficient between FWHM and the spin frequency derivative. The correlation coefficient changes randomly with time  implying that there is no obvious correlation between variation of spin-down rate and change of emission. \\

\acknowledgments \\
This work was supported by the National Key
R\&D Program of China (No. 2017YFA0402602),
 the National Natural Science Foundation of China
(11563008, 11873080, 11573059),
JY acknowledges supports from Center for Astronomical Mega-Science, Chinese Acade-\\my of Sciences.
RY is supported by the West Light Foundation of the Chinese Academy of Sciences, project 2016-QNXZ-B-24.
We are grateful to Feifei Kou for valuable suggestions, and Zhigang Wen for helpful discussion. 
The  Parkes  radio  telescope  is  part  of  the  Australia  Telescope  National  Facility  which  is  funded  by  the  Australian
Government for operation as a National Facility managed by
CSIRO.

%
%
%
%
\bibliographystyle{spr-mp-nameyear-cnd}
\bibliography{bibtex}

\begin{thebibliography}{42}
\ifx \bisbn   \undefined \def \bisbn  #1{ISBN #1}\fi
\ifx \binits  \undefined \def \binits#1{#1} \fi
\ifx \bauthor  \undefined \def \bauthor#1{#1} \fi
\ifx \batitle  \undefined \def \batitle#1{#1} \fi
\ifx \bjtitle  \undefined \def \bjtitle#1{#1}\fi
\ifx \bvolume  \undefined \def \bvolume#1{\textbf{#1}}\fi
\ifx \byear  \undefined \def \byear#1{#1} \fi
\ifx \bissue  \undefined \def \bissue#1{#1} \fi
\ifx \bfpage  \undefined \def \bfpage#1{#1} \fi
\ifx \blpage  \undefined \def \blpage #1{#1} \fi
\ifx \burl  \undefined \def \burl#1{\textsf{#1}} \fi
\ifx \doiurl  \undefined \def \doiurl#1{\textsf{#1}} \fi
\ifx \betal  \undefined \def \betal{\textit{et al.}} \fi
\ifx \binstitute  \undefined \def \binstitute#1{#1} \fi
\ifx \binstitutionaled  \undefined \def \binstitutionaled#1{#1} \fi
\ifx \bctitle  \undefined \def \bctitle#1{#1} \fi
\ifx \beditor  \undefined \def \beditor#1{#1} \fi
\ifx \bpublisher  \undefined \def \bpublisher#1{#1} \fi
\ifx \bbtitle  \undefined \def \bbtitle#1{#1} \fi
\ifx \bedition  \undefined \def \bedition#1{#1} \fi
\ifx \bseriesno  \undefined \def \bseriesno#1{#1} \fi
\ifx \blocation  \undefined \def \blocation#1{#1} \fi
\ifx \bsertitle  \undefined \def \bsertitle#1{#1} \fi
\ifx \bsnm \undefined \def \bsnm#1{#1} \fi
\ifx \bsuffix \undefined \def \bsuffix#1{#1} \fi
\ifx \bparticle \undefined \def \bparticle#1{#1} \fi
\ifx \barticle \undefined \def \barticle#1{#1} \fi
\ifx \bconfdate \undefined \def \bconfdate #1{#1} \fi
\ifx \botherref \undefined \def \botherref #1{#1} \fi
\ifx \url \undefined \def \url#1{\textsf{#1}} \fi
\ifx \bchapter \undefined \def \bchapter#1{#1} \fi
\ifx \bbook \undefined \def \bbook#1{#1} \fi
\ifx \bcomment \undefined \def \bcomment#1{#1} \fi
\ifx \oauthor \undefined \def \oauthor#1{#1} \fi
\ifx \citeauthoryear \undefined \def \citeauthoryear#1{#1} \fi
\ifx \endbibitem  \undefined \def \endbibitem {}\fi
\ifx \bconflocation  \undefined \def \bconflocation#1{#1} \fi
\ifx \arxivurl  \undefined \def \arxivurl#1{\textsf{#1}} \fi

\bibitem[\protect\citeauthoryear{{Backer}}{1970}]{Backer+1970}
\begin{barticle}
\bauthor{\bsnm{{Backer}}, \binits{D.C.}}:
\bjtitle{\nat}
\bvolume{228}(\bissue{5278}),
\bfpage{1297}
(\byear{1970})
\end{barticle}
\endbibitem

\bibitem[\protect\citeauthoryear{{Basu} and {Mitra}}{2018}]{Basu+Mitra+2018}
\begin{barticle}
\bauthor{\bsnm{{Basu}}, \binits{R.}},
\bauthor{\bsnm{{Mitra}}, \binits{D.}}:
\bjtitle{\mnras}
\bvolume{476}(\bissue{1}),
\bfpage{1345}
(\byear{2018}).
\arxivurl{1802.00315}
\end{barticle}
\endbibitem

\bibitem[\protect\citeauthoryear{{Bhattacharyya}
  et~al.}{2010}]{Bhattacharyya+etal+2010}
\begin{barticle}
\bauthor{\bsnm{{Bhattacharyya}}, \binits{B.}},
\bauthor{\bsnm{{Gupta}}, \binits{Y.}},
\bauthor{\bsnm{{Gil}}, \binits{J.}}:
\bjtitle{\mnras}
\bvolume{408}(\bissue{1}),
\bfpage{407}
(\byear{2010}).
\arxivurl{1006.0377}
\end{barticle}
\endbibitem

\bibitem[\protect\citeauthoryear{{Brook} et~al.}{2019}]{Brook+etal+2019}
\begin{barticle}
\bauthor{\bsnm{{Brook}}, \binits{P.R.}},
\bauthor{\bsnm{{Karastergiou}}, \binits{A.}},
\bauthor{\bsnm{{Johnston}}, \binits{S.}}:
\bjtitle{\mnras}
\bvolume{488}(\bissue{4}),
\bfpage{5702}
(\byear{2019}).
\arxivurl{1904.10989}
\end{barticle}
\endbibitem

\bibitem[\protect\citeauthoryear{{Brook} et~al.}{2016}]{Brook+etal+2016}
\begin{barticle}
\bauthor{\bsnm{{Brook}}, \binits{P.R.}},
\bauthor{\bsnm{{Karastergiou}}, \binits{A.}},
\bauthor{\bsnm{{Johnston}}, \binits{S.}},
\bauthor{\bsnm{{Kerr}}, \binits{M.}},
\bauthor{\bsnm{{Shannon}}, \binits{R.M.}},
\bauthor{\bsnm{{Roberts}}, \binits{S.J.}}:
\bjtitle{\mnras}
\bvolume{456}(\bissue{2}),
\bfpage{1374}
(\byear{2016}).
\arxivurl{1511.05481}
\end{barticle}
\endbibitem

\bibitem[\protect\citeauthoryear{{Brook} et~al.}{2018}]{Brook+etal+2018}
\begin{barticle}
\bauthor{\bsnm{{Brook}}, \binits{P.R.}},
\bauthor{\bsnm{{Karastergiou}}, \binits{A.}},
\bauthor{\bsnm{{McLaughlin}}, \binits{M.A.}},
\bauthor{\bsnm{{Lam}}, \binits{M.T.}},
\bauthor{\bsnm{{Arzoumanian}}, \binits{Z.}},
\bauthor{\bsnm{{Chatterjee}}, \binits{S.}},
\bauthor{\bsnm{{Cordes}}, \binits{J.M.}},
\bauthor{\bsnm{{Crowter}}, \binits{K.}},
\bauthor{\bsnm{{DeCesar}}, \binits{M.}},
\bauthor{\bsnm{{Demorest}}, \binits{P.B.}},
\bauthor{\bsnm{{Dolch}}, \binits{T.}},
\bauthor{\bsnm{{Ellis}}, \binits{J.A.}},
\bauthor{\bsnm{{Ferdman}}, \binits{R.D.}},
\bauthor{\bsnm{{Ferrara}}, \binits{E.}},
\bauthor{\bsnm{{Fonseca}}, \binits{E.}},
\bauthor{\bsnm{{Gentile}}, \binits{P.A.}},
\bauthor{\bsnm{{Jones}}, \binits{G.}},
\bauthor{\bsnm{{Jones}}, \binits{M.L.}},
\bauthor{\bsnm{{Lazio}}, \binits{T.J.W.}},
\bauthor{\bsnm{{Levin}}, \binits{L.}},
\bauthor{\bsnm{{Lorimer}}, \binits{D.R.}},
\bauthor{\bsnm{{Lynch}}, \binits{R.S.}},
\bauthor{\bsnm{{Ng}}, \binits{C.}},
\bauthor{\bsnm{{Nice}}, \binits{D.J.}},
\bauthor{\bsnm{{Pennucci}}, \binits{T.T.}},
\bauthor{\bsnm{{Ransom}}, \binits{S.M.}},
\bauthor{\bsnm{{Ray}}, \binits{P.S.}},
\bauthor{\bsnm{{Spiewak}}, \binits{R.}},
\bauthor{\bsnm{{Stairs}}, \binits{I.H.}},
\bauthor{\bsnm{{Stinebring}}, \binits{D.R.}},
\bauthor{\bsnm{{Stovall}}, \binits{K.}},
\bauthor{\bsnm{{Swiggum}}, \binits{J.K.}},
\bauthor{\bsnm{{Zhu}}, \binits{W.W.}}:
\bjtitle{\apj}
\bvolume{868}(\bissue{2}),
\bfpage{122}
(\byear{2018}).
\arxivurl{1810.08269}
\end{barticle}
\endbibitem

\bibitem[\protect\citeauthoryear{{Chen} et~al.}{2011}]{Chen+etal+2011}
\begin{barticle}
\bauthor{\bsnm{{Chen}}, \binits{J.L.}},
\bauthor{\bsnm{{Wang}}, \binits{H.G.}},
\bauthor{\bsnm{{Wang}}, \binits{N.}},
\bauthor{\bsnm{{Lyne}}, \binits{A.}},
\bauthor{\bsnm{{Liu}}, \binits{Z.Y.}},
\bauthor{\bsnm{{Jessner}}, \binits{A.}},
\bauthor{\bsnm{{Yuan}}, \binits{J.P.}},
\bauthor{\bsnm{{Kramer}}, \binits{M.}}:
\bjtitle{\apj}
\bvolume{741}(\bissue{1}),
\bfpage{48}
(\byear{2011}).
\arxivurl{1107.4676}
\end{barticle}
\endbibitem

\bibitem[\protect\citeauthoryear{{Cheng}}{1987a}]{cheng1987b}
\begin{barticle}
\bauthor{\bsnm{{Cheng}}, \binits{K.S.}}:
\bjtitle{\apj}
\bvolume{321},
\bfpage{805}
(\byear{1987}a)
\end{barticle}
\endbibitem

\bibitem[\protect\citeauthoryear{{Cheng}}{1987b}]{cheng1987a}
\begin{barticle}
\bauthor{\bsnm{{Cheng}}, \binits{K.S.}}:
\bjtitle{\apj}
\bvolume{321},
\bfpage{799}
(\byear{1987}b)
\end{barticle}
\endbibitem

\bibitem[\protect\citeauthoryear{{Coles} et~al.}{2011}]{Coles+etal+2011}
\begin{barticle}
\bauthor{\bsnm{{Coles}}, \binits{W.}},
\bauthor{\bsnm{{Hobbs}}, \binits{G.}},
\bauthor{\bsnm{{Champion}}, \binits{D.J.}},
\bauthor{\bsnm{{Manchester}}, \binits{R.N.}},
\bauthor{\bsnm{{Verbiest}}, \binits{J.P.W.}}:
\bjtitle{\mnras}
\bvolume{418}(\bissue{1}),
\bfpage{561}
(\byear{2011}).
\arxivurl{1107.5366}
\end{barticle}
\endbibitem

\bibitem[\protect\citeauthoryear{{Cordes}}{1979}]{Cordes+1979}
\begin{barticle}
\bauthor{\bsnm{{Cordes}}, \binits{J.M.}}:
\bjtitle{\ssr}
\bvolume{24}(\bissue{4}),
\bfpage{567}
(\byear{1979})
\end{barticle}
\endbibitem

\bibitem[\protect\citeauthoryear{{Cordes}}{1993}]{Cordes+James+1993}
\begin{bchapter}
\bauthor{\bsnm{{Cordes}}, \binits{J.M.}}:
In: \beditor{\bsnm{{Phillips}}, \binits{J.A.}},
\beditor{\bsnm{{Thorsett}}, \binits{S.E.}},
\beditor{\bsnm{{Kulkarni}}, \binits{S.R.}} (eds.)
\bbtitle{Planets Around Pulsars}.
\bsertitle{Astronomical Society of the Pacific Conference Series},
vol. \bseriesno{36},
p. \bfpage{43}
(\byear{1993})
\end{bchapter}
\endbibitem

\bibitem[\protect\citeauthoryear{{Desvignes}
  et~al.}{2019}]{Desvignes+etal+2019}
\begin{barticle}
\bauthor{\bsnm{{Desvignes}}, \binits{G.}},
\bauthor{\bsnm{{Kramer}}, \binits{M.}},
\bauthor{\bsnm{{Lee}}, \binits{K.}},
\bauthor{\bsnm{{van Leeuwen}}, \binits{J.}},
\bauthor{\bsnm{{Stairs}}, \binits{I.}},
\bauthor{\bsnm{{Jessner}}, \binits{A.}},
\bauthor{\bsnm{{Cognard}}, \binits{I.}},
\bauthor{\bsnm{{Kasian}}, \binits{L.}},
\bauthor{\bsnm{{Lyne}}, \binits{A.}},
\bauthor{\bsnm{{Stappers}}, \binits{B.W.}}:
\bjtitle{Science}
\bvolume{365}(\bissue{6457}),
\bfpage{1013}
(\byear{2019}).
\arxivurl{1909.06212}
\end{barticle}
\endbibitem

\bibitem[\protect\citeauthoryear{{Edwards}
  et~al.}{2006}]{Edwards+Hobbs+Manchester+2006}
\begin{barticle}
\bauthor{\bsnm{{Edwards}}, \binits{R.T.}},
\bauthor{\bsnm{{Hobbs}}, \binits{G.B.}},
\bauthor{\bsnm{{Manchester}}, \binits{R.N.}}:
\bjtitle{\mnras}
\bvolume{372}(\bissue{4}),
\bfpage{1549}
(\byear{2006}).
\arxivurl{astro-ph/0607664}
\end{barticle}
\endbibitem

\bibitem[\protect\citeauthoryear{{Folkner} et~al.}{2009}]{Folkner+etal+2009}
\begin{barticle}
\bauthor{\bsnm{{Folkner}}, \binits{W.M.}},
\bauthor{\bsnm{{Williams}}, \binits{J.G.}},
\bauthor{\bsnm{{Boggs}}, \binits{D.H.}}:
\bjtitle{Interplanetary Network Progress Report}
\bvolume{42-178},
\bfpage{1}
(\byear{2009})
\end{barticle}
\endbibitem

\bibitem[\protect\citeauthoryear{{Hobbs} et~al.}{2010}]{Hobbs+etal+2010}
\begin{barticle}
\bauthor{\bsnm{{Hobbs}}, \binits{G.}},
\bauthor{\bsnm{{Lyne}}, \binits{A.G.}},
\bauthor{\bsnm{{Kramer}}, \binits{M.}}:
\bjtitle{\mnras}
\bvolume{402}(\bissue{2}),
\bfpage{1027}
(\byear{2010}).
\arxivurl{0912.4537}
\end{barticle}
\endbibitem

\bibitem[\protect\citeauthoryear{{Hobbs}
  et~al.}{2006}]{Hobbs+Edwards+Manchester+2006}
\begin{barticle}
\bauthor{\bsnm{{Hobbs}}, \binits{G.B.}},
\bauthor{\bsnm{{Edwards}}, \binits{R.T.}},
\bauthor{\bsnm{{Manchester}}, \binits{R.N.}}:
\bjtitle{\mnras}
\bvolume{369}(\bissue{2}),
\bfpage{655}
(\byear{2006}).
\arxivurl{astro-ph/0603381}
\end{barticle}
\endbibitem

\bibitem[\protect\citeauthoryear{{Hotan} et~al.}{2004}]{Hotan+etal+2004}
\begin{barticle}
\bauthor{\bsnm{{Hotan}}, \binits{A.W.}},
\bauthor{\bsnm{{van Straten}}, \binits{W.}},
\bauthor{\bsnm{{Manchester}}, \binits{R.N.}}:
\bjtitle{\pasa}
\bvolume{21}(\bissue{3}),
\bfpage{302}
(\byear{2004}).
\arxivurl{astro-ph/0404549}
\end{barticle}
\endbibitem

\bibitem[\protect\citeauthoryear{{Johnston} et~al.}{1995}]{Johnston+etal+1995}
\begin{barticle}
\bauthor{\bsnm{{Johnston}}, \binits{S.}},
\bauthor{\bsnm{{Manchester}}, \binits{R.N.}},
\bauthor{\bsnm{{Lyne}}, \binits{A.G.}},
\bauthor{\bsnm{{Kaspi}}, \binits{V.M.}},
\bauthor{\bsnm{{D'Amico}}, \binits{N.}}:
\bjtitle{\aap}
\bvolume{293},
\bfpage{795}
(\byear{1995})
\end{barticle}
\endbibitem

\bibitem[\protect\citeauthoryear{{Johnston} et~al.}{1992}]{Johnston+etal+1992}
\begin{barticle}
\bauthor{\bsnm{{Johnston}}, \binits{S.}},
\bauthor{\bsnm{{Lyne}}, \binits{A.G.}},
\bauthor{\bsnm{{Manchester}}, \binits{R.N.}},
\bauthor{\bsnm{{Kniffen}}, \binits{D.A.}},
\bauthor{\bsnm{{D'Amico}}, \binits{N.}},
\bauthor{\bsnm{{Lim}}, \binits{J.}},
\bauthor{\bsnm{{Ashworth}}, \binits{M.}}:
\bjtitle{\mnras}
\bvolume{255},
\bfpage{401}
(\byear{1992})
\end{barticle}
\endbibitem

\bibitem[\protect\citeauthoryear{{Jones}}{1990}]{Jones+etal+1990}
\begin{barticle}
\bauthor{\bsnm{{Jones}}, \binits{P.B.}}:
\bjtitle{\mnras}
\bvolume{246},
\bfpage{364}
(\byear{1990})
\end{barticle}
\endbibitem

\bibitem[\protect\citeauthoryear{{Keith} et~al.}{2013}]{Keith+etal+2013}
\begin{barticle}
\bauthor{\bsnm{{Keith}}, \binits{M.J.}},
\bauthor{\bsnm{{Shannon}}, \binits{R.M.}},
\bauthor{\bsnm{{Johnston}}, \binits{S.}}:
\bjtitle{\mnras}
\bvolume{432}(\bissue{4}),
\bfpage{3080}
(\byear{2013}).
\arxivurl{1304.4644}
\end{barticle}
\endbibitem

\bibitem[\protect\citeauthoryear{{Kou} et~al.}{2018}]{Kou+etal+2018}
\begin{barticle}
\bauthor{\bsnm{{Kou}}, \binits{F.F.}},
\bauthor{\bsnm{{Yuan}}, \binits{J.P.}},
\bauthor{\bsnm{{Wang}}, \binits{N.}},
\bauthor{\bsnm{{Yan}}, \binits{W.M.}},
\bauthor{\bsnm{{Dang}}, \binits{S.J.}}:
\bjtitle{\mnras}
\bvolume{478}(\bissue{1}),
\bfpage{24}
(\byear{2018}).
\arxivurl{1801.01248}
\end{barticle}
\endbibitem

\bibitem[\protect\citeauthoryear{{Kramer} et~al.}{2006}]{kramer+2006}
\begin{barticle}
\bauthor{\bsnm{{Kramer}}, \binits{M.}},
\bauthor{\bsnm{{Lyne}}, \binits{A.G.}},
\bauthor{\bsnm{{O'Brien}}, \binits{J.T.}},
\bauthor{\bsnm{{Jordan}}, \binits{C.A.}},
\bauthor{\bsnm{{Lorimer}}, \binits{D.R.}}:
\bjtitle{Science}
\bvolume{312}(\bissue{5773}),
\bfpage{549}
(\byear{2006}).
\arxivurl{astro-ph/0604605}
\end{barticle}
\endbibitem

\bibitem[\protect\citeauthoryear{{Li} et~al.}{2016}]{Li+etal+2016}
\begin{barticle}
\bauthor{\bsnm{{Li}}, \binits{L.}},
\bauthor{\bsnm{{Wang}}, \binits{N.}},
\bauthor{\bsnm{{Yuan}}, \binits{J.P.}},
\bauthor{\bsnm{{Wang}}, \binits{J.B.}},
\bauthor{\bsnm{{Hobbs}}, \binits{G.}},
\bauthor{\bsnm{{Lentati}}, \binits{L.}},
\bauthor{\bsnm{{Manchester}}, \binits{R.N.}}:
\bjtitle{\mnras}
\bvolume{460}(\bissue{4}),
\bfpage{4011}
(\byear{2016}).
\arxivurl{1605.08574}
\end{barticle}
\endbibitem

\bibitem[\protect\citeauthoryear{{Lyne} et~al.}{2010}]{Lyne+etal+2010}
\begin{barticle}
\bauthor{\bsnm{{Lyne}}, \binits{A.}},
\bauthor{\bsnm{{Hobbs}}, \binits{G.}},
\bauthor{\bsnm{{Kramer}}, \binits{M.}},
\bauthor{\bsnm{{Stairs}}, \binits{I.}},
\bauthor{\bsnm{{Stappers}}, \binits{B.}}:
\bjtitle{Science}
\bvolume{329}(\bissue{5990}),
\bfpage{408}
(\byear{2010}).
\arxivurl{1006.5184}
\end{barticle}
\endbibitem

\bibitem[\protect\citeauthoryear{{Lyubarskii}}{1996}]{Lyubarskii+1996}
\begin{barticle}
\bauthor{\bsnm{{Lyubarskii}}, \binits{Y.E.}}:
\bjtitle{\aap}
\bvolume{311},
\bfpage{172}
(\byear{1996})
\end{barticle}
\endbibitem

\bibitem[\protect\citeauthoryear{{Melatos} and {Link}}{2014}]{melatos2014}
\begin{barticle}
\bauthor{\bsnm{{Melatos}}, \binits{A.}},
\bauthor{\bsnm{{Link}}, \binits{B.}}:
\bjtitle{\mnras}
\bvolume{437}(\bissue{1}),
\bfpage{21}
(\byear{2014}).
\arxivurl{1310.3108}
\end{barticle}
\endbibitem

\bibitem[\protect\citeauthoryear{{Parthasarathy}
  et~al.}{2019}]{Parthasarathy+etal+2019}
\begin{barticle}
\bauthor{\bsnm{{Parthasarathy}}, \binits{A.}},
\bauthor{\bsnm{{Shannon}}, \binits{R.M.}},
\bauthor{\bsnm{{Johnston}}, \binits{S.}},
\bauthor{\bsnm{{Lentati}}, \binits{L.}},
\bauthor{\bsnm{{Bailes}}, \binits{M.}},
\bauthor{\bsnm{{Dai}}, \binits{S.}},
\bauthor{\bsnm{{Kerr}}, \binits{M.}},
\bauthor{\bsnm{{Manchester}}, \binits{R.N.}},
\bauthor{\bsnm{{Os{\l}owski}}, \binits{S.}},
\bauthor{\bsnm{{Sobey}}, \binits{C.}},
\bauthor{\bsnm{{van Straten}}, \binits{W.}},
\bauthor{\bsnm{{Weltevrede}}, \binits{P.}}:
\bjtitle{\mnras}
\bvolume{489}(\bissue{3}),
\bfpage{3810}
(\byear{2019}).
\arxivurl{1908.11709}
\end{barticle}
\endbibitem

\bibitem[\protect\citeauthoryear{{Perera} et~al.}{2015}]{Perera+etal+2015}
\begin{barticle}
\bauthor{\bsnm{{Perera}}, \binits{B.B.P.}},
\bauthor{\bsnm{{Stappers}}, \binits{B.W.}},
\bauthor{\bsnm{{Weltevrede}}, \binits{P.}},
\bauthor{\bsnm{{Lyne}}, \binits{A.G.}},
\bauthor{\bsnm{{Bassa}}, \binits{C.G.}}:
\bjtitle{\mnras}
\bvolume{446}(\bissue{2}),
\bfpage{1380}
(\byear{2015}).
\arxivurl{1410.5228}
\end{barticle}
\endbibitem

\bibitem[\protect\citeauthoryear{{Perera} et~al.}{2016}]{Perera+etal+2016}
\begin{barticle}
\bauthor{\bsnm{{Perera}}, \binits{B.B.P.}},
\bauthor{\bsnm{{Stappers}}, \binits{B.W.}},
\bauthor{\bsnm{{Weltevrede}}, \binits{P.}},
\bauthor{\bsnm{{Lyne}}, \binits{A.G.}},
\bauthor{\bsnm{{Rankin}}, \binits{J.M.}}:
\bjtitle{\mnras}
\bvolume{455}(\bissue{1}),
\bfpage{1071}
(\byear{2016}).
\arxivurl{1510.04484}
\end{barticle}
\endbibitem

\bibitem[\protect\citeauthoryear{{Petroff} et~al.}{2013}]{Petroff+etal+2013}
\begin{barticle}
\bauthor{\bsnm{{Petroff}}, \binits{E.}},
\bauthor{\bsnm{{Keith}}, \binits{M.J.}},
\bauthor{\bsnm{{Johnston}}, \binits{S.}},
\bauthor{\bsnm{{van Straten}}, \binits{W.}},
\bauthor{\bsnm{{Shannon}}, \binits{R.M.}}:
\bjtitle{\mnras}
\bvolume{435}(\bissue{2}),
\bfpage{1610}
(\byear{2013}).
\arxivurl{1307.7221}
\end{barticle}
\endbibitem

\bibitem[\protect\citeauthoryear{{Rankin} et~al.}{2006}]{Rankin+etal+2006}
\begin{barticle}
\bauthor{\bsnm{{Rankin}}, \binits{J.M.}},
\bauthor{\bsnm{{Rodriguez}}, \binits{C.}},
\bauthor{\bsnm{{Wright}}, \binits{G.A.E.}}:
\bjtitle{\mnras}
\bvolume{370}(\bissue{2}),
\bfpage{673}
(\byear{2006}).
\arxivurl{astro-ph/0601368}
\end{barticle}
\endbibitem

\bibitem[\protect\citeauthoryear{{Rankin} et~al.}{2013}]{Rankin+etal+2013}
\begin{barticle}
\bauthor{\bsnm{{Rankin}}, \binits{J.M.}},
\bauthor{\bsnm{{Wright}}, \binits{G.A.E.}},
\bauthor{\bsnm{{Brown}}, \binits{A.M.}}:
\bjtitle{\mnras}
\bvolume{433}(\bissue{1}),
\bfpage{445}
(\byear{2013}).
\arxivurl{1304.7697}
\end{barticle}
\endbibitem

\bibitem[\protect\citeauthoryear{{Smits} et~al.}{2005}]{Smits+etal+2005}
\begin{barticle}
\bauthor{\bsnm{{Smits}}, \binits{J.M.}},
\bauthor{\bsnm{{Mitra}}, \binits{D.}},
\bauthor{\bsnm{{Kuijpers}}, \binits{J.}}:
\bjtitle{\aap}
\bvolume{440}(\bissue{2}),
\bfpage{683}
(\byear{2005}).
\arxivurl{astro-ph/0506264}
\end{barticle}
\endbibitem

\bibitem[\protect\citeauthoryear{{Stairs}
  et~al.}{2000}]{Stairs+Lyne+Shemar+2000}
\begin{barticle}
\bauthor{\bsnm{{Stairs}}, \binits{I.H.}},
\bauthor{\bsnm{{Lyne}}, \binits{A.G.}},
\bauthor{\bsnm{{Shemar}}, \binits{S.L.}}:
\bjtitle{\nat}
\bvolume{406}(\bissue{6795}),
\bfpage{484}
(\byear{2000})
\end{barticle}
\endbibitem

\bibitem[\protect\citeauthoryear{{Stairs} et~al.}{2019}]{Stairs+etal+2019}
\begin{barticle}
\bauthor{\bsnm{{Stairs}}, \binits{I.H.}},
\bauthor{\bsnm{{Lyne}}, \binits{A.G.}},
\bauthor{\bsnm{{Kramer}}, \binits{M.}},
\bauthor{\bsnm{{Stappers}}, \binits{B.W.}},
\bauthor{\bsnm{{van Leeuwen}}, \binits{J.}},
\bauthor{\bsnm{{Tung}}, \binits{A.}},
\bauthor{\bsnm{{Manchester}}, \binits{R.N.}},
\bauthor{\bsnm{{Hobbs}}, \binits{G.B.}},
\bauthor{\bsnm{{Lorimer}}, \binits{D.R.}},
\bauthor{\bsnm{{Melatos}}, \binits{A.}}:
\bjtitle{\mnras}
\bvolume{485}(\bissue{3}),
\bfpage{3230}
(\byear{2019}).
\arxivurl{1903.01573}
\end{barticle}
\endbibitem

\bibitem[\protect\citeauthoryear{{Taylor} and
  {Manchester}}{1975}]{Taylor+Manchester+1975}
\begin{barticle}
\bauthor{\bsnm{{Taylor}}, \binits{J.H.}},
\bauthor{\bsnm{{Manchester}}, \binits{R.N.}}:
\bjtitle{\aj}
\bvolume{80},
\bfpage{794}
(\byear{1975})
\end{barticle}
\endbibitem

\bibitem[\protect\citeauthoryear{{van Straten} and
  {Bailes}}{2011}]{Van+Bailes+2011}
\begin{barticle}
\bauthor{\bsnm{{van Straten}}, \binits{W.}},
\bauthor{\bsnm{{Bailes}}, \binits{M.}}:
\bjtitle{\pasa}
\bvolume{28}(\bissue{1}),
\bfpage{1}
(\byear{2011}).
\arxivurl{1008.3973}
\end{barticle}
\endbibitem

\bibitem[\protect\citeauthoryear{{Wang} et~al.}{2007}]{Wang+etal+2007}
\begin{barticle}
\bauthor{\bsnm{{Wang}}, \binits{N.}},
\bauthor{\bsnm{{Manchester}}, \binits{R.N.}},
\bauthor{\bsnm{{Johnston}}, \binits{S.}}:
\bjtitle{\mnras}
\bvolume{377}(\bissue{3}),
\bfpage{1383}
(\byear{2007}).
\arxivurl{astro-ph/0703241}
\end{barticle}
\endbibitem

\bibitem[\protect\citeauthoryear{{Wang} et~al.}{2000}]{Wang+etal+2000}
\begin{barticle}
\bauthor{\bsnm{{Wang}}, \binits{N.}},
\bauthor{\bsnm{{Manchester}}, \binits{R.N.}},
\bauthor{\bsnm{{Pace}}, \binits{R.T.}},
\bauthor{\bsnm{{Bailes}}, \binits{M.}},
\bauthor{\bsnm{{Kaspi}}, \binits{V.M.}},
\bauthor{\bsnm{{Stappers}}, \binits{B.W.}},
\bauthor{\bsnm{{Lyne}}, \binits{A.G.}}:
\bjtitle{\mnras}
\bvolume{317}(\bissue{4}),
\bfpage{843}
(\byear{2000}).
\arxivurl{astro-ph/0005561}
\end{barticle}
\endbibitem

\bibitem[\protect\citeauthoryear{{Wang} et~al.}{2001}]{wang+etal+2001}
\begin{barticle}
\bauthor{\bsnm{{Wang}}, \binits{N.}},
\bauthor{\bsnm{{Manchester}}, \binits{R.N.}},
\bauthor{\bsnm{{Zhang}}, \binits{J.}},
\bauthor{\bsnm{{Wu}}, \binits{X.J.}},
\bauthor{\bsnm{{Yusup}}, \binits{A.}},
\bauthor{\bsnm{{Lyne}}, \binits{A.G.}},
\bauthor{\bsnm{{Cheng}}, \binits{K.S.}},
\bauthor{\bsnm{{Chen}}, \binits{M.Z.}}:
\bjtitle{\mnras}
\bvolume{328}(\bissue{3}),
\bfpage{855}
(\byear{2001}).
\arxivurl{astro-ph/0111006}
\end{barticle}
\endbibitem

\end{thebibliography}


\begin{thebibliography}{}
\end{thebibliography}

\end{document}